\documentclass[iop]{emulateapj}
 \usepackage{amsmath, amsthm, amssymb,amsfonts}
\usepackage{graphicx}
\usepackage{dcolumn}
\usepackage{bm}
\usepackage{color}
\usepackage[abs]{overpic}

\begin{document}

\title{Evolution of switchbacks in the inner Heliosphere}

\author{Anna Tenerani}
\affiliation{Department of Physics, University of Texas at Austin, TX 78712}
  \email{Anna.Tenerani@austin.utexas.edu}
  \author{Nikos Sioulas}%
   \affiliation{Department of Earth, Planetary, and Space Sciences, UCLA, Los Angeles, CA, 90095}
   \author{Lorenzo Matteini}
 \affiliation{Imperial College, London}
    \author{Olga Panasenco}
 \affiliation{Advanced Heliophysics}
 \author{Chen Shi}%
 \affiliation{Department of Earth, Planetary, and Space Sciences, UCLA, Los Angeles, CA, 90095}
\author{Marco Velli}%
 \affiliation{Department of Earth, Planetary, and Space Sciences, UCLA, Los Angeles, CA, 90095}

\date{\today}

\begin{abstract}
We analyze magnetic field data from the first six encounters of PSP, three Helios fast streams and two Ulysses south polar passes covering heliocentric distances $0.1\lesssim R\lesssim 3$~au. We use this data set to statistically determine the evolution of switchbacks of different periods and amplitudes with distance from the Sun. We compare the radial evolution of  magnetic field variances with that of the mean square amplitudes of switchbacks, and quantify the radial evolution of the cumulative counts of switchbacks per km. We find that the amplitudes of switchbacks decrease faster than the overall turbulent fluctuations, in a way consistent with the radial decrease of the mean magnetic field. This could be the result of a saturation of amplitudes and may be a signature of decay processes of large amplitude Alfv\'enic fluctuations in the solar wind. We find that the evolution of switchback occurrence in the solar wind is  scale-dependent: the fraction of longer duration switchbacks increases with radial distance  whereas it decreases for shorter switchbacks. This implies that switchback dynamics is a complex process involving both decay and in-situ generation in the inner heliosphere. We confirm that switchbacks can be generated by the expansion although other type of switchbacks generated closer to the sun cannot be ruled out.     
\end{abstract}

\maketitle

\section{Introduction} 

We have analyzed data from Parker Solar Probe (PSP), Helios and Ulysses to investigate the evolution of switchbacks with heliocentric distance ranging from about 0.1 to 3 au.  Switchbacks are embedded within a continuous flux of turbulent, Alfv\'enic fluctuations that permeates the solar wind and that dominates the frequency range $f\simeq[10^{-6}-10^{-1}]$~Hz of the magnetic field energy spectrum. Switchbacks make up the subset of largely Alfv\'enic fluctuations whose amplitude is large enough that the magnetic field kinks  backwards on itself, leading to a local field polarity reversal and to a corresponding radial velocity jet  \citep{matteini_2014}.  Although switchbacks have been observed in the past, both by Helios at distances $R\simeq 0.3-1$~au from the sun \citep{horbury_MNRAS}, and by Ulysses beyond 1~au, one of the major findings of PSP  is that this peculiar type of fluctuation/structure appears to be ubiquitous and a prominent feature of the solar wind closer to the sun, at distances   $R\lesssim 0.2$~au. Understanding what is the origin of switchbacks remains an important and open issue because it may exclude or point to  mechanisms taking place in the corona related to coronal heating and solar wind acceleration. They may also contribute to feeding the turbulent cascade via their nonlinear evolution.

Many  mechanisms have  been proposed to explain switchbacks. It has been suggested that they are remnants of processes in the  corona such as interchange reconnection \citep{drake_2020, zank_2020}. In a high beta plasma, they may also originate from  instabilities like the firehose \citep{tenerani_2018}, which is known to lead to highly kinked field lines. Switchbacks may also form dynamically as turbulent fluctuations propagate into the inner heliosphere, induced by shear flows or by the solar wind expansion itself \citep{landi_2006,ruffolo_2020,squire_2020,mallet_2021,schwadron_2021}.

Switchbacks have now been characterized extensively in their  local properties \citep{horbury_2020, woolley_2020, woodham_2021, laker_2021}. They are three dimensional field-aligned structures displaying the typical velocity-magnetic field correlation that characterizes Alfv\'en waves propagating away from the sun. Just like the overall  turbulent, Alfv\'enic fluctuations, switchbacks are characterized by a nearly constant magnetic field magnitude, a condition corresponding to spherical polarization \citep{tsurutani_2018, matteini_2015}. A small  level of compressibility has been observed by PSP in some switchbacks \citep{farrell_2020,larosa_2021}, especially at the boundaries, a property that may be related to their evolutionary stage but that still remains to be investigated.  

Interestingly,  large amplitude, Alfv\'enic fluctuations (in the sense of velocity-magnetic field correlation) and switchbacks, with constant total magnetic field magnitude, provide an exact, nonlinear dynamical state at the magnetohydrodynamic scales \citep{goldstein_1974, barnes_1974, hollweg_1974}. However, such a nonlinear state is thought to be unstable, since large amplitude Alfv\'enic fluctuations tend to decay due to resonant couplings with compressible modes via parametric and modulational instabilities, proven to persist in various conditions, including expansion effects and realistic temperature anisotropies. Theoretical investigations and numerical simulations support the idea that Alfv\'enic fluctuations of coronal origins should  already begin to decay close to the sun, say, within tens of solar radii \citep{tenerani_2013, reville_2018, shoda_2018}.  Recently, we have shown via  numerical simulations that a localized switchback is more stable than an Alfv\'enic wave of non constant total field magnitude (like the one considered by, e.g., \citet{landi_2005}), or of a  periodic circularly or arc polarized shear Alfv\'en wave \citep{delzanna_2001}, and that it can propagate for distances up to a few tens of solar radii \citep{tenerani_2020}. However, parametric decay eventually sets in, destroying the switchback  well before it can travel distances of the order of 1~au. This suggests that in the absence of ongoing dynamical forcing -- such as solar wind expansion, for example -- capable of continually replenishing switchbacks in-situ, their occurrence  rate should decrease with radial distance. Vice-versa, the effect of a continuous forcing would lead to an increase  in the occurrence rate (or to a steady state) as we move further away from the sun. Understanding how fluctuations and switchbacks evolve with radial distance is relevant to understand not just how switchbacks can form out of the overall turbulent fluctuations, but also to understand how they may contribute to the evolution of the turbulent cascade  in the solar wind. 

In this paper,  we study the radial evolution of switchbacks in the inner heliosphere out to 3 au by combining data from PSP, Helios and Ulysses. We track the radial evolution of the magnetic energy density carried by switchbacks as compared to the overall turbulent fluctuations and we investigate the occurrence rate of switchbacks. We finally discuss how our results compare and fit into existing theories and observations of  switchback generation.   

\section{Data sets and methods}
In this work, we use magnetic field data from Helios 1 and 2 at 6 second resolution (in SSE coordinates; \citet{helios_mag}) and consider a subset of the  intervals analyzed by \citet{perrone_2019}, who classify data into three fast Alfv\'enic streams crossed by Helios at various radial distances during years 1975-1976. In this paper we will adopt their same notation to denote intervals pertaining to stream A, stream B and stream C, respectively. 

For Ulysses,  magnetic field data from the VHM instrument are resampled at 6 seconds resolution (RTN; \citet{ulysses_mag}). We consider the first and third south polar passes, that occurred in 1994 and 2006-2007 denoted PP94 and PP06, respectively, to compare data from different solar cycles. 

Finally,  we use high cadence magnetic field data from the FIELDS instrument (in RTN coordinates; \citet{bale_2016}) onboard Parker Solar Probe (PSP) resampled at 6 seconds during the first six orbits (E1 through E6), from year 2018 to year 2020. The same analysis was performed with 1-second resolution and results were qualitatively similar. The PSP intervals have been identified by visual inspection to avoid current sheet crossings (leading to north-south main polarity reversals) and to select only those intervals with small values of magnetic compressibility  $\delta |{\bf B}|/|{\bf B}|<0.2$, on 5-minute averages, although the majority of intervals had even lower values, $\delta |{\bf B}|/|{\bf B}|\lesssim 0.1$. For the perihelia of E1, E2, E4 and E6 we have also determined what intervals correspond to the same source region via PFSS extrapolation \citep{pan20}, in order to compare switchbacks counts within the same or nearby streams. Plasma data from Helios \citep{helios_mag}, SWOOPS \citep{swoops}, and SPC \citep{spc} are also used to infer the mean radial speed of the wind for each time interval. 

Mean magnetic field, variances and mean amplitude of switchbacks, are obtained via temporal averages  over a sliding time window of length $\tau$ with $\tau\in[30$~min $,12$~h], hence we consider time scales of fluctuations that fall well into the inertial range and energy-containing range of Alfv\'enic turbulence. For a fixed value of the time $\tau$ we calculate the mean magnetic field $<{\bf B}>_\tau$  and the variances of the magnetic field components $<\delta {B_i}^2>_\tau$.  Switchbacks are identified as follows. For each time scale $\tau$ we search for those intervals in which the angle $\theta$ between the total magnetic field  ${\bf B}$ and  $<{\bf B}>_{\tau}$ is larger than $\theta=90^\circ$. In this way we only select fluctuations that correspond to a backwards kink  of the magnetic field. By using the direction of the total mean magnetic field rather than the radial direction to identify switchbacks allows us to avoid including  small radial fluctuations in a non-dominantly radial mean magnetic field. The amplitude of the switchbacks in the identified  interval is  defined as $(\delta B_i)_{\tau}^2=E[(B_i-<B_i>_\tau)^2]$, where $E[\cdot]$ stands for the mean, that in the case of switchbacks is evaluated inside the field reversal. This procedure is also used to determine the cumulative counts of switchbacks, essentially the number of switchbacks of duration less than $\tau$. We adopt the same method to determine the duration of the field reversal $\delta t$, and we use this information to investigate the evolution of the probability distribution (pdf) of switchback duration with radial distance. Finally, for each data set of  $<\delta {B_i}^2>_\tau$ and $(\delta B_i)_{\tau}^2$ we then take the mean to obtain the radial trends of such quantities. Count rates of switchbacks at given $\tau$ are obtained by normalizing the number of switchbacks events in a given interval of duration $\Delta t$  by the distance $\Delta R$ spanned in that interval by the switchbacks, which is $\Delta R=\Delta t |V_{sw}-V_{sp}|$ with $V_{sp}$ the average speed of the spacecraft, neglecting the contribution from the Alfv\'en speed (we verified on PSP data that its inclusion does not change our results).  
 

\section{Results}

\subsection{Radial evolution of the energy density of fluctuations}

We have analyzed the radial evolution of fluctuations amplitude from the Helios data within each of the three fast streams, during the polar  passes of Ulysses, and by combining PSP data. Data from some Helios streams were also used in the past by \citet{villante_1982} and \citet{bavassano_1982} to infer the radial evolution of turbulence properties. In particular, they showed that the rms amplitude of magnetic field fluctuations obey the WKB prediction $\delta b_{rms}\simeq R^{-3/2}$ to a good approximation at the lowest frequencies ($f\simeq 10^{-4}-10^{-3}$~Hz), whereas at higher frequencies they decay faster with radial distance due to the ongoing nonlinear cascade. Our results are consistent with such findings, and similar trends are also found at PSP and Ulysses.  

\begin{figure}[htb]
 \includegraphics[width=0.45\textwidth]{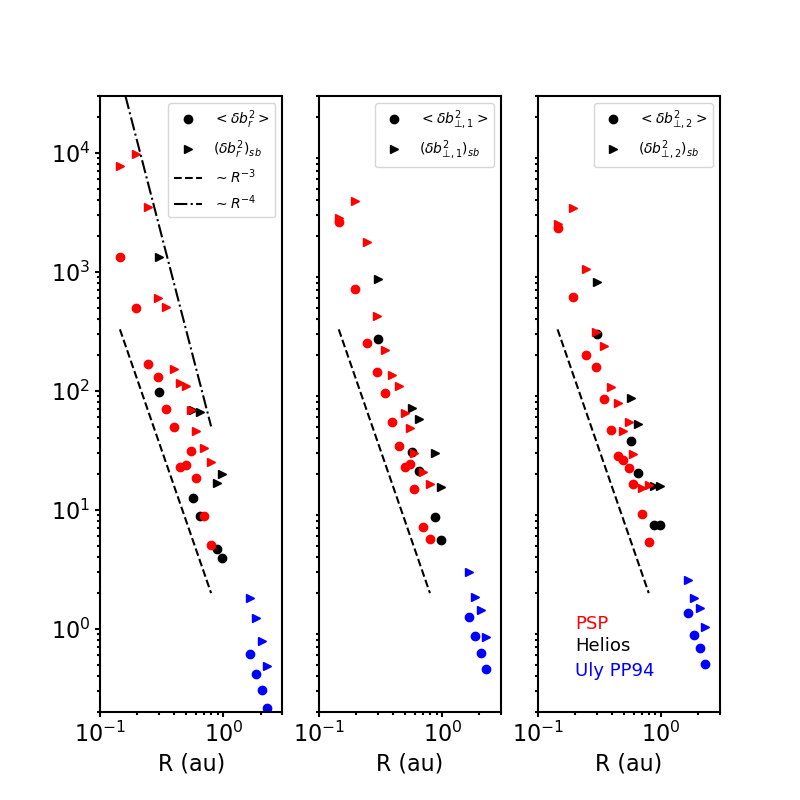}
\caption{Variance (dots) and mean square amplitude of switchbacks (triangles) as a function of radial distance for the radial (left panel), and perpendicular (middle and right panels) components of the magnetic field obtained for $\tau=8$~h. The dashed line shows the scaling $\sim R^{-3}$ and the dot-dashed $R^{-4}$.}
\label{rms}
\end{figure}

Figure~\ref{rms} shows an example of the radial evolution of the variances of the magnetic field for each component  $<\delta B_i^2>$ and of the mean square amplitude $(\delta B_{i})^2$  of the identified switchbacks for stream A of Helios (in black color), PSP (in red) and Ulysses PP94 (in blue) for $\tau=8$~h. PSP data points  from PSP (43 in total) have been binned in intervals of width $\Delta R=0.05$~au, and the dashed lines represent  reference slopes of $\sim R^{-3}$ for the variances, and $\sim R^{-4}$ for the mean square amplitude of switchbacks.The three panels from left to right show the radial and two perpendicular components of the field.

\begin{figure}[htb]
 \includegraphics[width=0.45\textwidth]{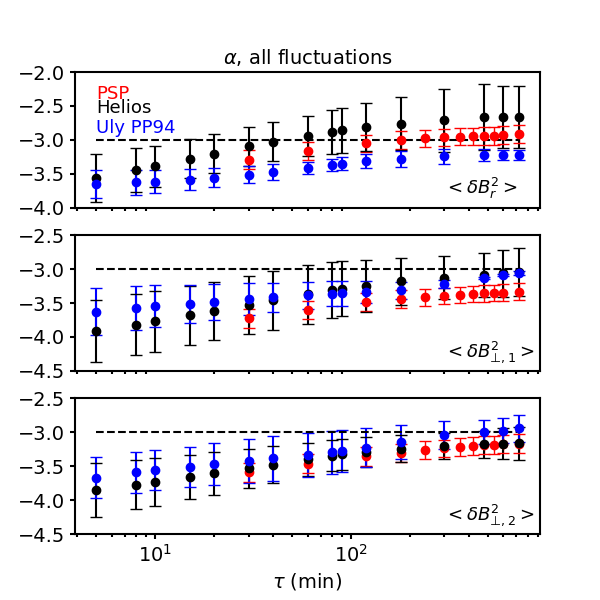}
\caption{Angular coefficient $\alpha$ at different $\tau$ obtained from the best fit  of magnetic field variances as a function of radial distance in log-log scale. The dashed horizontal lines mark the value $\alpha=-3$.}
\label{alpha_rms}
\end{figure}

\begin{figure}[htb]
 \includegraphics[width=0.45\textwidth]{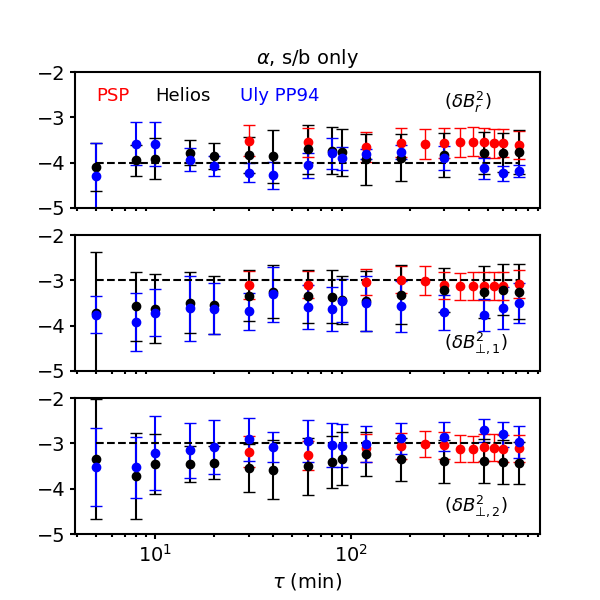}
\caption{Same as Fig.~\ref{alpha_rms} but for the mean square amplitude of switchbacks only. The dashed line in the top panel marks the value $\alpha=-4$.}
\label{alpha_sb}
\end{figure}

By assuming a log-log scale linear relation of amplitudes with radial distance $\log \delta B^2=A+\alpha \log R$, where $R$ is in units of au, we have performed a best fit analysis of the variances of the magnetic field and of the mean square amplitude of switchbacks at various values of $\tau$. Results obtained from PSP E1-E6, Ulysses PP94 and from the average of the three Helios streams A, B and C are displayed in Fig.~\ref{alpha_rms} and Fig.~\ref{alpha_sb}. Fig.~\ref{alpha_rms} shows that variances follow the radial trend predicted by the WKB theory to a good approximation, with $\alpha\lesssim -3$ for the tangential and normal components (and $y,z$ in SSE coordinates) and $\alpha \gtrsim -3$ for the radial component. While the coefficients $\alpha$ for the non-radial components are consistent among the three different spacecraft, we notice that the radial component of the fluctuations decays at different rates in Ulysses with respect to PSP and Helios. In particular the difference between the obtained angular coefficient $\alpha$ from Ulysses and Helios (or PSP) is larger than the estimated error, pointing to the fact that at 1 au and beyond, other mechanisms, such as nonlinearities and possibly interactions with large scale shear flows, affect the evolution of fluctuations also at the larger scales.  In general, however, we recover the trends found by \citep{villante_1982, bavassano_1982}, with the fluctuation energy at shorter timescales decaying with radial distance faster than the larger timescales. The coefficient $\alpha$  displays a monotonic dependence with $\tau$ and a change of slope at around $\tau_*\simeq 10^2$~min can be seen in all the three components, and at the three spacecraft. The timescale $\tau_*\simeq 10^2$ min roughly corresponds to the spectral break that separates the inertial range from the $f^{-1}$ range of the turbulent spectrum. 
This particular time scale depends on radial distance, however, we had to consider a wide range of heliocentric distances in order to find the best fit for the angular coefficient $\alpha$. As a consequence,  we cannot capture clearly the shift of $\tau_*$ with $R$.  Fig.~\ref{alpha_sb} shows the same quantities but for the mean square amplitude of switchbacks only. Contrary to the overall variances, the  coefficient $\alpha$ does not depend strongly on $\tau$. The mean values of $\alpha$ for the radial, tangential ($y$ for SSE) and normal ($z$ for SSE) are listed in Table~\ref{alpha_av}. For comparison, in the table we report also the mean $\alpha$ values for the variances obtained for values $\tau<100$ min (left values) and $\tau>100$ min (right values). As can be seen, the radial component of switchbacks decays much faster than the overall turbulent spectrum with a coefficient of about $\alpha\simeq -4$, while the non-radial components have a decay rate $-4<\alpha<-3$ which is consistent with the decay rate of transverse turbulent fluctuations. This suggests that  switchbacks behave as a separate population within the turbulent flux of Alfv\'enic fluctuations in the sense that their energy density overall decays faster, pointing to an ongoing continuous dynamical evolution of these structures. 

\begin{table*}[htp]
\caption{Mean value of the  power-law coefficient $\alpha$ for switchbacks (first three columns) and of the variances (last three columns). The mean value of $\alpha$ for the variances has been evaluated for $\tau<100$ min (left values) and $\tau>100$ min (right values).}
\begin{center}
\begin{tabular}{c c c c c c c}
\hline
\hline
 &$<\alpha_r>_{sb}$ & $<\alpha_{\bot,1}>_{sb}$ & $<\alpha_{\bot,2}>_{sb}$ & $<\alpha_r>$& $<\alpha_{\bot,1}>$ &  $<\alpha_{\bot,2}>$\\
PSP E1-E6 & 	$-3.6$ &  $ -3.1$	& $-3.1$  & $-3.2, -3.0$	& $-3.7, -3.4$	& $-3.5, -3.2$	\\
Helios   &	$-3.8$ & $-3.4$	& $-3.4$ & $-3.2, -2.7$ & $-3.6, -3.1$  & $-3.6, -3.2$\\
Uly PP94   &	$-3.9$ & $-3.6$& $-3.0$	& $-3.5, -3.3$ 	& $-3.5, -3.2$	& $-3.5, -3.1$\\
\hline
\hline
\end{tabular}
\end{center}
\label{alpha_av}
\end{table*}%

\subsection{Cumulative counts of switchbacks/km and pdf}
\begin{figure}[htb]
 \includegraphics[width=0.4\textwidth]{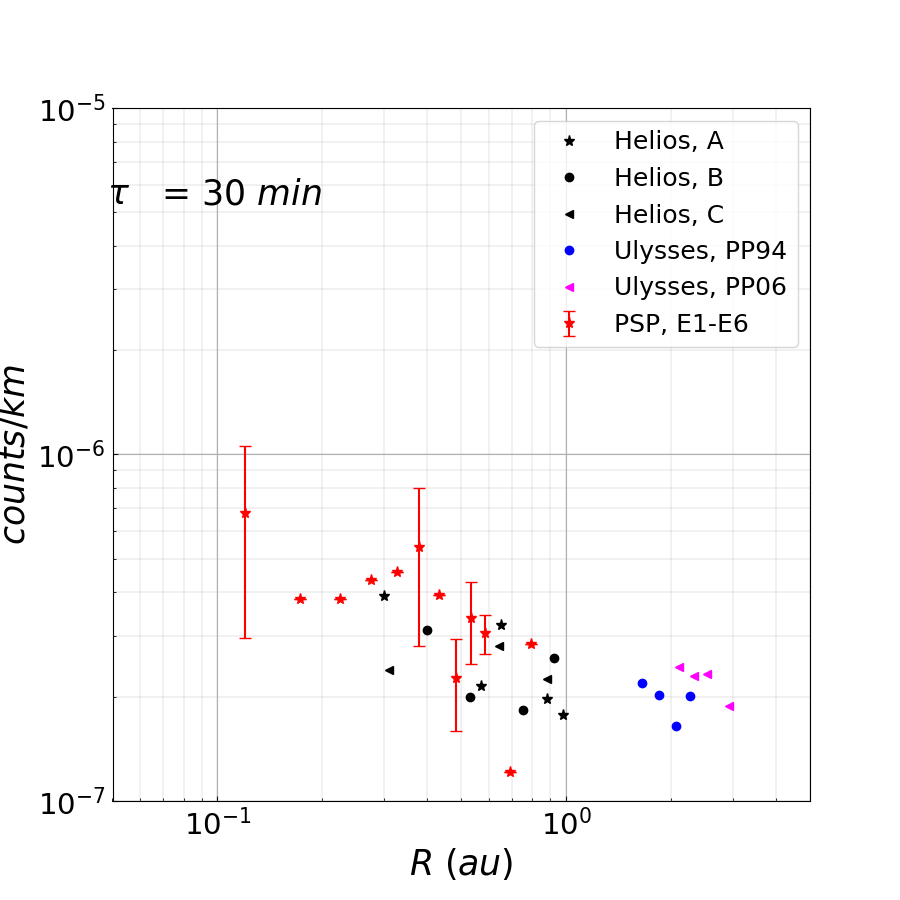}
  \includegraphics[width=0.4\textwidth]{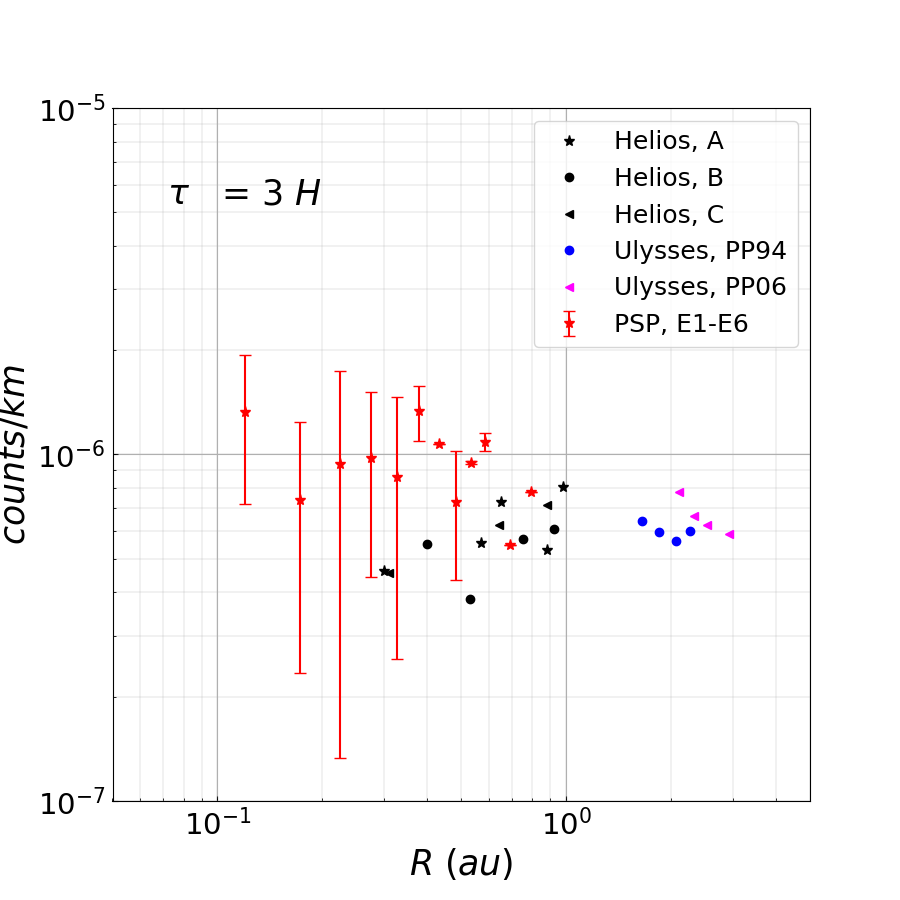}
    \includegraphics[width=0.4\textwidth]{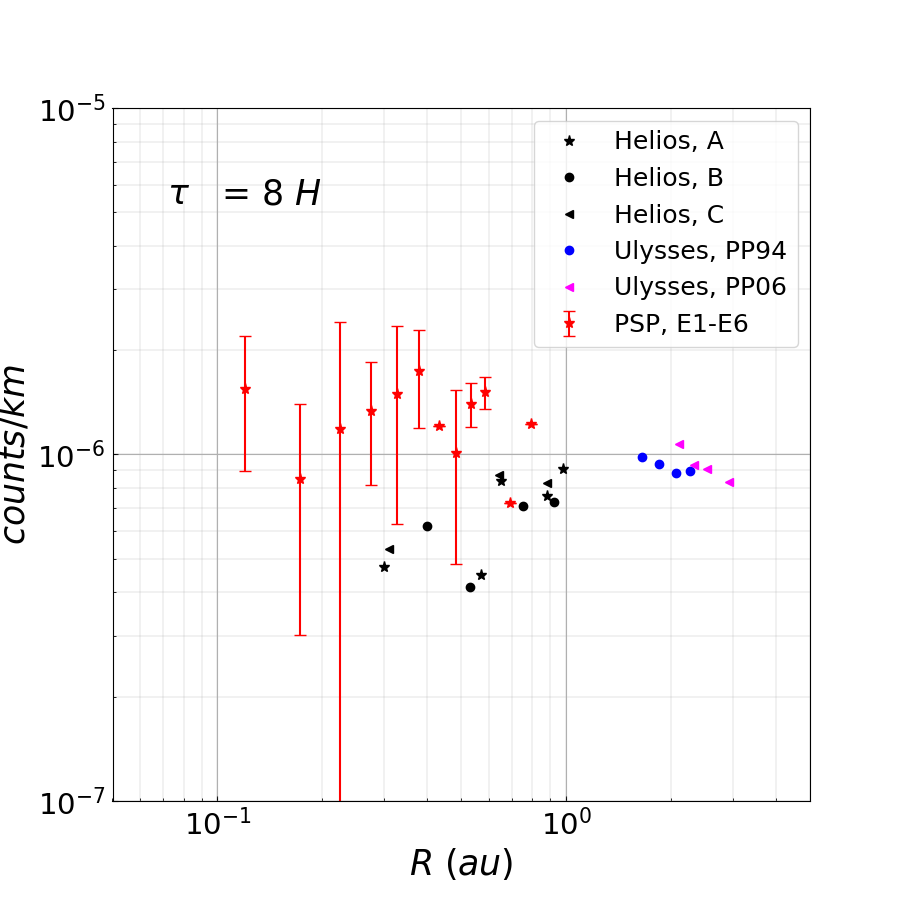}
\caption{Cumulative counts of switchbacks per $km$ obtained for three different values of $\tau$.}
\label{rates_all}
\end{figure}

Figure \ref{rates_all} shows the cumulative counts of switchbacks per km as a function of $R$ at  different values of $\tau$. We have chosen three representative values, $\tau=30$~min, 3~h and 8~h, to show how trends change as longer switchbacks are considered. Data points from PSP (48 in total) have been binned inside intervals of width $\Delta R=0.05$~au, and the error bars represent the range of points within each bin. Count rates from stream A, B, C, and PP94, PP06 are much less scattered than PSP  and on average they lie systematically below  PSP rates as $\tau$ increases above 1~h. Count rates for $\tau=30$~min in both PSP and Helios show a decreasing rate with radial distance, although PSP data points are quite scattered. As timescale increases, and thus  switchbacks of longer duration are included, Helios displays a net radial increase in the count rates  while  PSP rates have a more oscillatory behavior.  At Ulysses, count rates appear to somewhat decrease with radial distance regardless of the scale considered. 
This suggests that switchbacks undergo a complex evolution, where expansion or other in-situ effects are efficient in inducing magnetic field folds at larger scales while others decay, especially the shorter ones. We notice that such trends are clear particularly in Helios, where switchbacks can be followed at several radial distances within a given stream. We argue that the combined effect of the presence of decaying and forming switchbacks and the mixing of different source regions and streams can explain the highly scattered data points found with PSP. For this reason we have also calculated the cumulative counts per km for E1, E2, E4 and E6 by considering only those intervals that correspond to streams originating from nearby regions on the sun (not shown here). However, due to the few points satisfying such conditions, it is difficult to infer solid trends for the count rates in PSP. We find indeed that some streams display an increasing rate of switchbacks with radial distance while other a decreasing trend (not shown).  Further away from the sun, within the relatively steady fast streams observed by Ulysses, effects of the magnetic spiral may also affect the count rates because the non-radial components of the mean magnetic field decrease slower than fluctuations amplitude. At Ulysses, the tangential mean magnetic field can reach up to half the mean radial magnetic field. On average, the mean tangential field forms an angle of about $20^\circ$ with the radial direction. This deviation from a purely radial mean magnetic field may be sufficient to affect the strength of the fluctuation amplitude relative to the mean magnetic field, inhibiting the formation of switchbacks by expansion effects (see discussion).

\begin{figure}[htb]
 \includegraphics[width=0.37\textwidth]{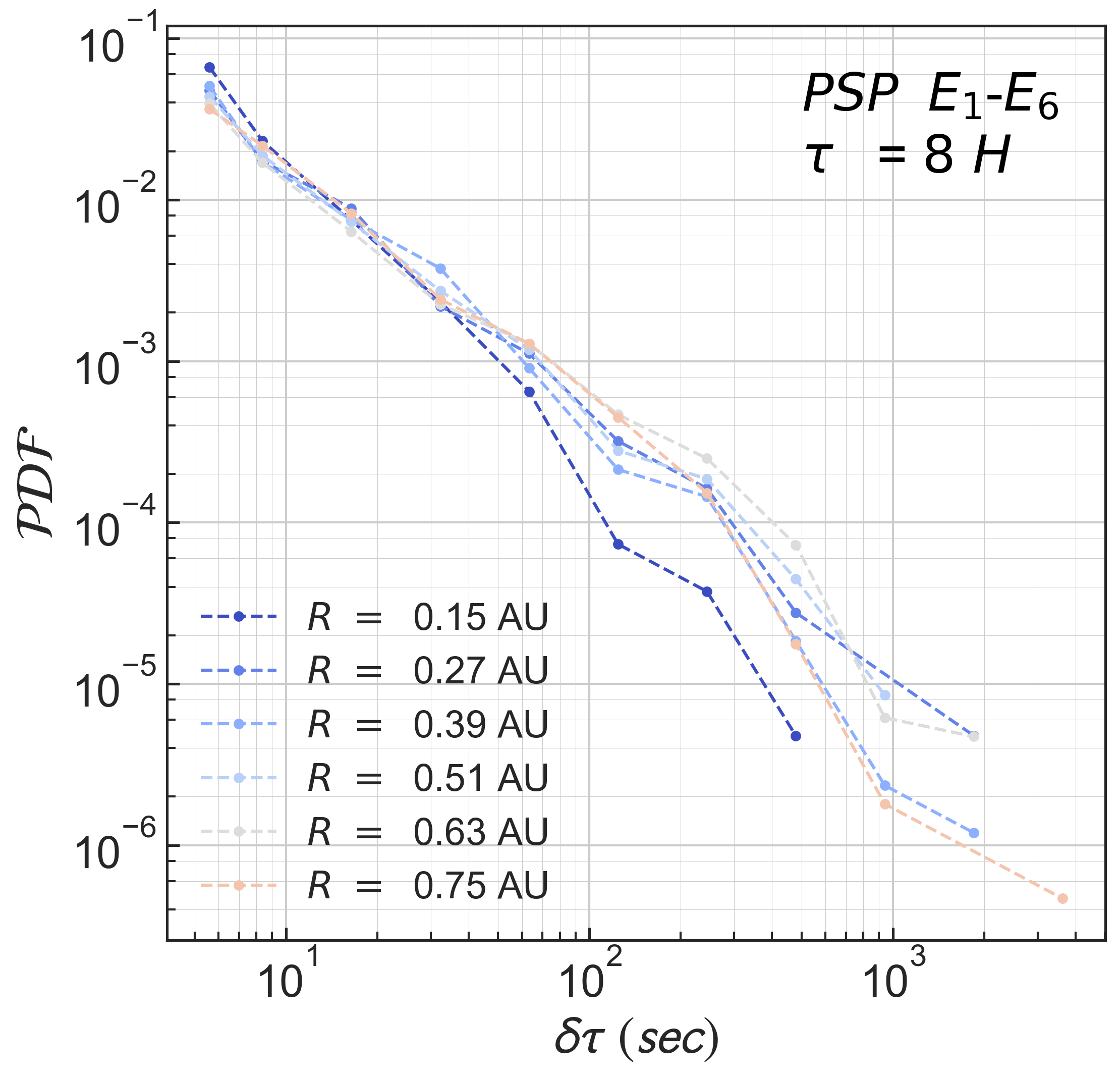}
  \includegraphics[width=0.4\textwidth]{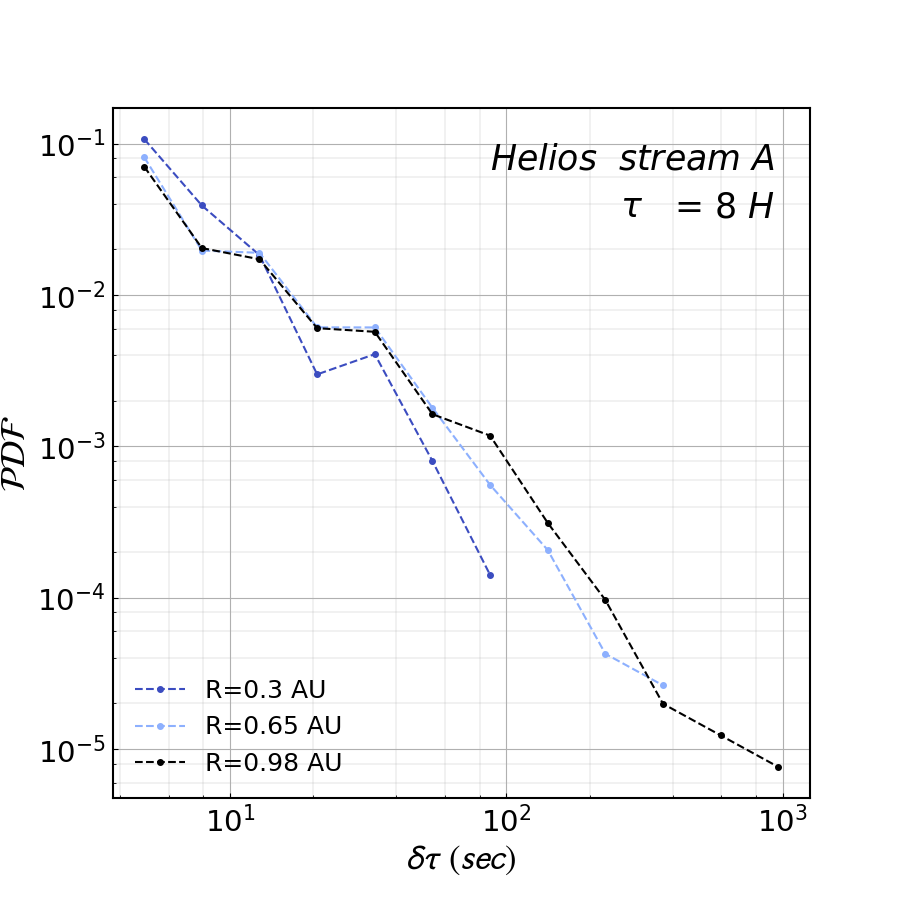}
    \includegraphics[width=0.4\textwidth]{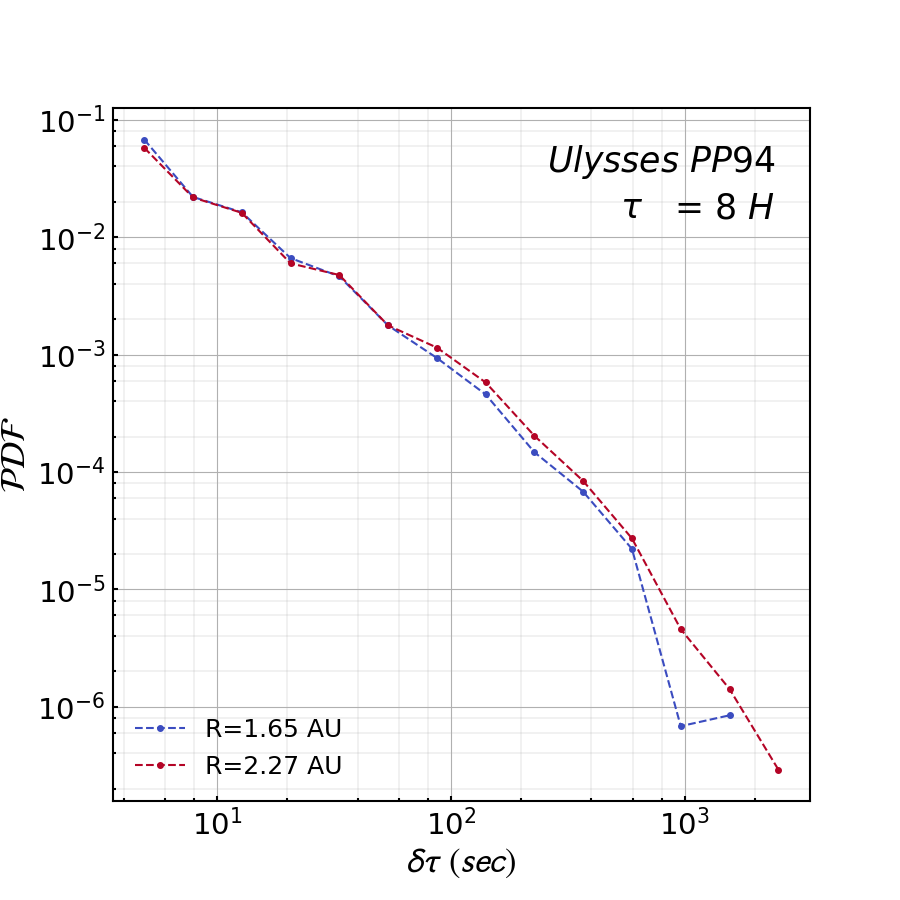}
\caption{Probability distribution function of switchback duration $\delta t$ for $\tau=8$~h.}
\label{pdfs}
\end{figure}

In order to gain  insight on the evolution of switchbacks with radial distance,  in Fig.~\ref{pdfs} we  show the probability distribution of switchbacks as a function of their duration   $\delta t$.   In this case we have chosen a relatively long value of $\tau$, $\tau=8$~h, to include a wide range of time scales. Pdfs for larger values of $\tau$ (not shown here) display the same qualitative features. Since count rates for PSP are very scattered, a finer radial resolution of the pdf provides a better representation of its radial evolution.  For PSP data we therefore show the pdfs for 6 mean radial distances. For Helios and Ulysses, instead, we show the pdfs at the closest and furthest heliocentric distance for stream A and PP94, respectively (other Helios streams and PP06 are similar). Inspection of the pdfs shows that shorter duration switchbacks are more common than the longer ones, in agreement with previous work \citep{dudok_2020}.  However, our analysis also shows that the fraction of switchbacks of longer duration  increases with radial distance, while the opposite trend is typically seen at the shorter scales, for $\delta t\approx 10-100$~s.

\section{Discussion}
A well known property of shear Alfv\'en waves in the expanding solar wind is that, for frequencies higher than the expansion rate, they evolve according to the conservation law of wave action \citep{bretherton_1969, jacques_ApJ_1977}.   In the radially expanding solar wind 
this leads to fluctuation amplitudes scaling as $\delta B_i^2\propto R^{-3}$. Such a scaling holds true for any Alfv\'enic structure at arbitrary amplitude, involving also radial fluctuations, provided the magnitude of the total magnetic field remains constant, so that nonlinearities and coupling with compressible modes are quenched.  On the other hand,  this condition implies that  the tip of the magnetic field is bound to rotate on the surface of a sphere of radius $B$, where $B$ is the magnitude of the total magnetic field, thus imposing a constraint on the maximum possible excursion of the magnetic field \citep{matteini_2018}. 

Our data analysis shows that at the larger scales, those roughly lying in the $f^{-1}$ range of frequencies, the overall turbulent fluctuations follow to a good approximation the WKB trend in all three components of the magnetic field, thus, including the radial component. However, switchbacks behave as a separate population in that the radial component of their fluctuations  decreases with radial distance as the mean magnetic field,  $\delta B_r^2\sim R^{-4}$. In other words, while the expansion can naturally drive relative large amplitudes  for a radial mean magnetic field ($\delta B_r/<B_r>\propto R^{1/2}$) out from the turbulent bath of fluctuations, some other physical processes must come into play to prevent relative amplitudes from increasing without bound and maintaining the observed nearly constant-$B$ constraint. Compressible effects leading to parametric instabilities and wave steepening/collapse become dynamically important at large amplitudes of about $10\%$ the mean magnetic field, or more. Coupling with compressible modes may provide a possible way to control fluctuations amplitudes whereas kinetic effects have been shown to play an important role in controlling the level of plasma and magnetic compressibility  \citep{tenerani_2013,gonzalez_2021}. Our finding of a faster decay of the radial amplitudes may be an indication of such type of processes, although a focused numerical study on this problem needs to be done.  Recent  simulations in the expanding solar wind have indeed shown that switchbacks can form in-situ \citep{squire_2020, shoda_2021}, however a study of how switchbacks evolve once they are generated still needs to be carried out and it will be of interest to compare numerical results with our observations.

\citet{mozer_2021} have performed a study of the occurrence rate of switchbacks using data from  encounters 3 through 7 of PSP. The authors adopt a Poisson regression model to infer the dependence of the occurrence rate (counts per hour) of switchbacks on solar wind speed and radial distance, and conclude that the occurrence rate depends on the wind speed (higher rates for higher wind speed)  and does not depend on radial distance, thus, they exclude in-situ generation mechanisms. However, our analysis, which includes the wind speed in the switchbacks count normalization, shows that switchbacks counts with PSP are extremely scattered and it is difficult to infer a dominant trend with radial distance. The highly scattered occurrence of switchbacks is likely due to the mixing of different streams, as also mentioned by \citet{mozer_2021}, but also to the presence of switchbacks that decay and reform in the wind.  Our analysis indeed also shows that the occurrence of switchbacks is scale-dependent, a trend particularly clear in the cumulative counts in the Helios streams, and also shown explicitly in the pdfs from PSP, Helios and Ulysses data. Specifically, we have found that the fraction of switchbacks of duration of a few tens of seconds and longer increases with radial distance and that the fraction of those of duration below a few tens of seconds instead decreases. Switchbacks in the solar wind can decay and reform in the expanding solar wind, with in-situ generation being more efficient at the larger scales. 

\section{Summary}
We have analyzed magnetic field data from PSP, Helios and Ulysses covering heliocentric distances $0.15\lesssim R\lesssim 3$~au at 6 second resolution. We have determined the radial evolution of the amplitudes of turbulent fluctuations by comparing the magnetic field variances with the mean square amplitudes of switchbacks, and quantified the radial evolution of the cumulative counts of switchbacks per km. Our results can be summarized as follows: (1) radial amplitudes of switchbacks ($\delta B_r$) decrease faster than that of the overall turbulent fluctuations. Specifically, their amplitude square scales approximately as $\delta B_r^2\sim R^{-4}$ at all scales as opposed to the bath of turbulent fluctuations scaling between $R^{-3}$  at the larger scales (in agreement with the WKB prediction), and faster at the shorter scales.  Thus, switchbacks amplitudes saturate  with respect to the radial mean field; (2) the evolution of Alfv\'enic fluctuations at larger scales (obeying WKB trends) in the expanding wind is a likely driver for in-situ switchback generation; (3) the occurrence of switchbacks in the solar wind depends on the duration of the switchbacks. Both in-situ formation and decay are at play in the solar wind; (4) in-situ switchback generation is more efficient at larger scales, however, we observe a net decrease of the cumulative counts per km of switchbacks beyond 1 au at Ulysses, regardless of their duration. We conclude that it is possible that switchbacks of two or more type can coexist: those generated close to the sun that decay as they propagate away, and those that are reformed or maintained alive in the inner heliosphere by a combination of expansion effects (or other in-situ drivers) and decay processes. At increasing distances $R\gtrsim 1$ au, effects of the Parker spiral and possibly nonlinearities and large scale shears may be the cause of the observed net decrease of switchbacks cumulative counts per km.  
  
\acknowledgements{This research was supported by NASA grant \#80NSS\-C18K1211 and the HERMES NASA DRIVE Science Center Grant \#80NSSC20K0604.}

\end{document}